\documentclass[english]{revtex4-2}
\usepackage[LGR,T1]{fontenc}
\usepackage[latin9]{inputenc}
\usepackage{geometry}
\usepackage{babel}
\usepackage{array}
\usepackage{refstyle}
\usepackage{float}
\usepackage{booktabs}
\usepackage{units}
\usepackage{amstext}
\usepackage{graphicx}
\usepackage{setspace}
\usepackage[unicode=true,pdfusetitle,
 bookmarks=true,bookmarksnumbered=false,bookmarksopen=false,
 breaklinks=false,pdfborder={0 0 0},pdfborderstyle={},backref=false,colorlinks=false]
 {hyperref}

\makeatletter

\AtBeginDocument{\providecommand\Figref[1]{\ref{Fig:#1}}}
\AtBeginDocument{\providecommand\eqref[1]{\ref{eq:#1}}}
\AtBeginDocument{\providecommand\Tabref[1]{\ref{Tab:#1}}}
\AtBeginDocument{\providecommand\figref[1]{\ref{fig:#1}}}
\DeclareRobustCommand{\greektext}{%
  \fontencoding{LGR}\selectfont\def\encodingdefault{LGR}}
\DeclareRobustCommand{\textgreek}[1]{\leavevmode{\greektext #1}}
\ProvideTextCommand{\~}{LGR}[1]{\char126#1}

\newcommand{\lyxmathsym}[1]{\ifmmode\begingroup\def\b@ld{bold}
  \text{\ifx\math@version\b@ld\bfseries\fi#1}\endgroup\else#1\fi}

\providecommand{\tabularnewline}{\\}
\RS@ifundefined{subsecref}
  {\newref{subsec}{name = \RSsectxt}}
  {}
\RS@ifundefined{thmref}
  {\def\RSthmtxt{theorem~}\newref{thm}{name = \RSthmtxt}}
  {}
\RS@ifundefined{lemref}
  {\def\RSlemtxt{lemma~}\newref{lem}{name = \RSlemtxt}}
  {}

\usepackage{xcolor}
\usepackage{colortbl}
\hypersetup{
    colorlinks=false,
    linkbordercolor={lightgray},
	citebordercolor={lightgray},
}
\date{\today}

\@ifundefined{showcaptionsetup}{}{%
 \PassOptionsToPackage{caption=false}{subfig}}
\usepackage{subfig}
\makeatother

\begin{document}
\title{\noindent Modeling solute-grain boundary interactions in a bcc Ti-Mo
alloy using density functional theory}
\author{Hariharan Umashankar$^{a}$, Daniel Scheiber$^{b}$, Vsevolod I. Razumovskiy$^{b}$,
Matthias Militzer$^{a}$}
\address{$^{a}$Centre for Metallurgical Process Engineering, The University
of British Columbia, Vancouver, BC V6T 1Z4 Canada~\\
$^{b}$Materials Center Leoben Forschung GmbH, Roseggerstrasse 12,
Leoben 8700, Austria}
\begin{abstract}
Solute segregation in alloys is a key phenomenon which affects various material characteristics such as embrittlement, grain growth and precipitation kinetics. In this work, the segregation energies of Y, Zr, and Nb
to a \textgreek{S}5 grain boundary in a bcc Ti-25 at \% Mo alloy were
determined using density functional theory (DFT) calculations. A systematic
approach was laid out by computing the solution energy distributions
in the bulk alloy using Warren-Cowley short-range order parameters
to find a representative bulk-solute reference energy. Additionally,
different scenarios were considered when a solute atom replaces different
sites in terms of their local Ti-Mo chemistry at the GB plane to calculate
the distribution of segregation energies. The solute segregation to
a Mo site at the GB plane is preferred rather than to a Ti site. Further
analysis shows that these segregation energy trends can be rationalized
based on a primarily elastic interaction. Thus the segregation energies
scale with the solute size such that Y has the largest segregation
energies followed by Zr and Nb. \\

\end{abstract}
\maketitle

\section{Introduction}

%\begin{doublespace}
Titanium and it's alloys are commercially important materials and
have in particular applications in biomedical and aerospace industries
\citep{Banerjee2013Feb}. While many allotropes of Titanium exists,
there are three important ones: the $\alpha$-phase (hcp) is stable
at low temperature and atmospheric pressure, the $\beta$-phase (bcc)
is stable at high temperature and atmospheric pressure and the $\omega$-phase
(hexagonal) is stable at high pressures \citep{Zhang2008Oct}. Apart
from temperature and pressure, various alloying elements are added
in order to achieve stability of the different phases to arrive at
a desirable microstructure including dual-phase $\alpha$- $\beta$
structures \citep{Okazaki1993}. Solute elements may significantly
affect the microstructure evolution and the mechanical properties
of Ti alloys \citep{Niinomi1998Mar}. This is in part attributed to
segregation to grain boundaries (GBs) and other interfaces. Recent
advances in atomic-scale characterization using atom probe tomography
have made it possible to study solute segregation at interfaces in
many metallic alloy systems \citep{Ebner2021Dec,Gault2021Jul}. On
the other hand, atomistic simulations provide a detailed and efficient
way to analyze the segregation behavior of various solute elements
\citep{Mishin2010Feb}. First-principles simulations in Ti have been
limited to the $\alpha$-phase mainly focusing on segregation of non-metallic
interstitial elements and transition metals \citep{Aksyonov2017Sep,Hui2022Jan}
to twin and symmetric tilt GBs. First-principles studies of segregation
in the $\beta$-phase are challenging because this phase is thermodynamically
and mechanically unstable at 0 K.
%\end{doublespace}

%\begin{doublespace}
There are three approaches to stabilize $\beta$-Ti using first-principles
studies: one is to perform \emph{ab initio} molecular dynamics (AIMD)
at high-temperatures \citep{Korbmacher2019Sep}, second is alloying
with a $\beta$-phase stabilizing element such as Mo and V \citep{Marker2018Feb}
and third is to apply pressure to the system \citep{Hu2008Aug}. Due
to the exorbitant computationally expensive nature of AIMD calculations
and the un-physical nature of pressure for evaluating solute segregation,
the alloying strategy in $\beta$-Ti is considered for predicting
solute segregation to grain boundaries. Raabe \emph{et al.} \citep{Raabe2007Aug}
showed that 25 at.\% Mo is required to thermodynamically and mechanically
stabilize the bcc phase over the hcp phase. Yan \emph{et al.} \citep{Yan2017Jul}
have also used alloying as a tool to investigate the segregation of
Y and B in Ti-Mo and Ti-V alloys. They employed the virtual crystal
approximation (VCA) method where each atomic site is a representative
alloy with the overall composition. Their work shows that Y tends
to segregate while B tends to de-segregate at the $\Sigma3${[}110{]}/(111)
GB. 

Investigating defect structures and energetics in alloy systems using
first-principles calculations has been a recent field of study. An
emerging body of literature is available on evaluating point-defect
energetics such as vacancy formation energies in binary and multi-component
alloys \citep{Ruban2016Apr,Luo2018Feb,Esfandiarpour2019Sep,Zhang2021Apr,Zhang2022Apr,Zhou2022Aug}
whereas similar simulations for planar defects in alloys are still
in their evolving state. Shen \emph{et al. }\citep{Shen2021Mar} have
investigated the GB formation energies as a function of Mo and Nb
alloying elements in bcc $\gamma$-U. Their work shows that \textgreek{S}3{[}110{]}/(111)
and \textgreek{S}5{[}100{]}/(013)grain boundary structures are stable
at greater than 27 at.\% Mo and 30 at.\% Nb. Scheiber \emph{et al.}
\citep{Scheiber2015Apr} studied Re segregation and work of separation
for various <110> symmetric tilt grain boundaries in W-25 at.\% Re
alloys. Their studies show that segregation of Re increases the work
of separation which is consistent with experimental observations in
W-Re alloys. These simulations were limited to a few configurations
of the alloy structure and solute substitution in the host metal. 

We study in this work the segregation of a solute species in a binary alloy. Using density functional theory (DFT) calculations, the binary Ti-25at.\%Mo alloy is represented with the special quasi-random structure (SQS) approach and all possible configurations with the solute are included when evaluating the energetics of the bulk and GB structure. Representative values are computed and analyzed for bulk
solution energies and segregation energies of Y, Zr and Nb to the
$\Sigma$5(130){[}001{]} GB in a Ti- 25 at.\% Mo alloy. Further, segregation
of the host elements (Ti and Mo) to the GB plane are calculated and
finally the computed energies are analysed in terms of the local volume
and chemistry.
%\end{doublespace}

\section{Computational method}

\subsection{Structural models for bulk and GBs}

%\begin{doublespace}
\noindent Ti-25 at.\% Mo is chosen as the alloy composition for our
calculations. The random alloy is modelled within the SQS super-cell approach wherein the alloy constituents are distributed randomly within the cell. For creating the SQS alloy structures, the Warren-Cowley short range order (SROs) parameter \citep{Cowley1950Mar,Cowley1965May}
is computed for six consecutive nearest neighbor (NN) shells to ensure low short range order within the simulation cells. A structure
with SRO=0 for first NN and SRO < 0.01 for the other five shells is
considered as a representative Ti-Mo alloy structure. For the bulk
structure, a 128 atom supercell is used with dimensions of $12.85\times12.85\times12.85$
\AA$^3$ as shown in \Figref{bulk-supercell-128}. In order to
obtain energies of a Ti and Mo atom in a Ti- 25at.\% Mo structure,
a 125 atom supercell is also created (which contains 93 Ti with 31Mo
+ 1Mo), i.e., with a slightly higher (25.6 at.\% ) Mo content. The
SROs of the 125 atom supercell is set to be similar to that of the
128-atom supercell. For creating the GB structure, first a 60-atom
bulk structure is generated with similar SROs as the 128-atom supercell
with x, y, z axes aligned as {[}130{]}, {[}$\bar{3}$10{]} and {[}001{]}
crystallographic directions respectively. A $\Sigma$5(130){[}001{]}
GB bcc Ti-25 at.\% Mo structure with dimensions of $10.16\times20.32\times9.64$
\AA$^3$, containing 120 atoms is constructed by choosing a y-plane
to mirror across in 60 atom supercell. Ten different alloy GB structures
are possible to create with this approach. Out of the ten, two GB
structures are selected to study solute segregation. These structures
are highlighted in \Figref{gb-1} and \Figref{gb-2}. 
%\end{doublespace}
\begin{center}
\begin{figure}[H]
\begin{centering}
\subfloat[\label{fig:bulk-supercell-128}]{\begin{centering}
\includegraphics[scale=0.33]{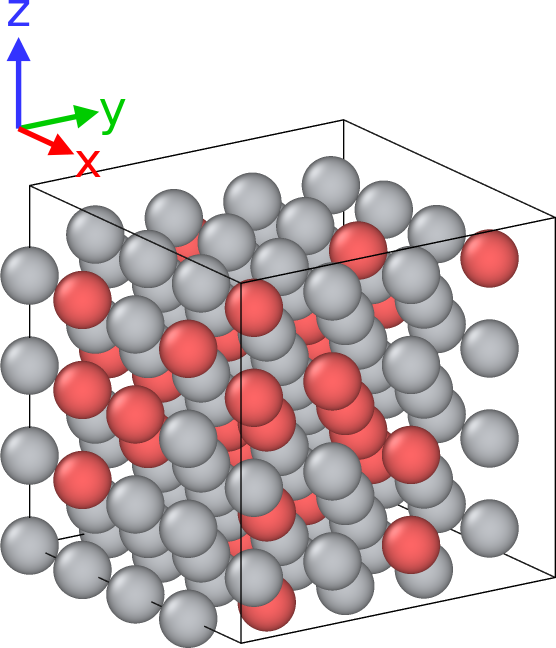}\includegraphics[scale=0.1]{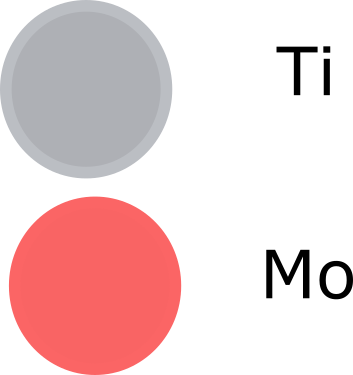}
\par\end{centering}
}
\par\end{centering}
\begin{centering}
\subfloat[\label{fig:gb-1}]{\centering{}\includegraphics[scale=0.3]{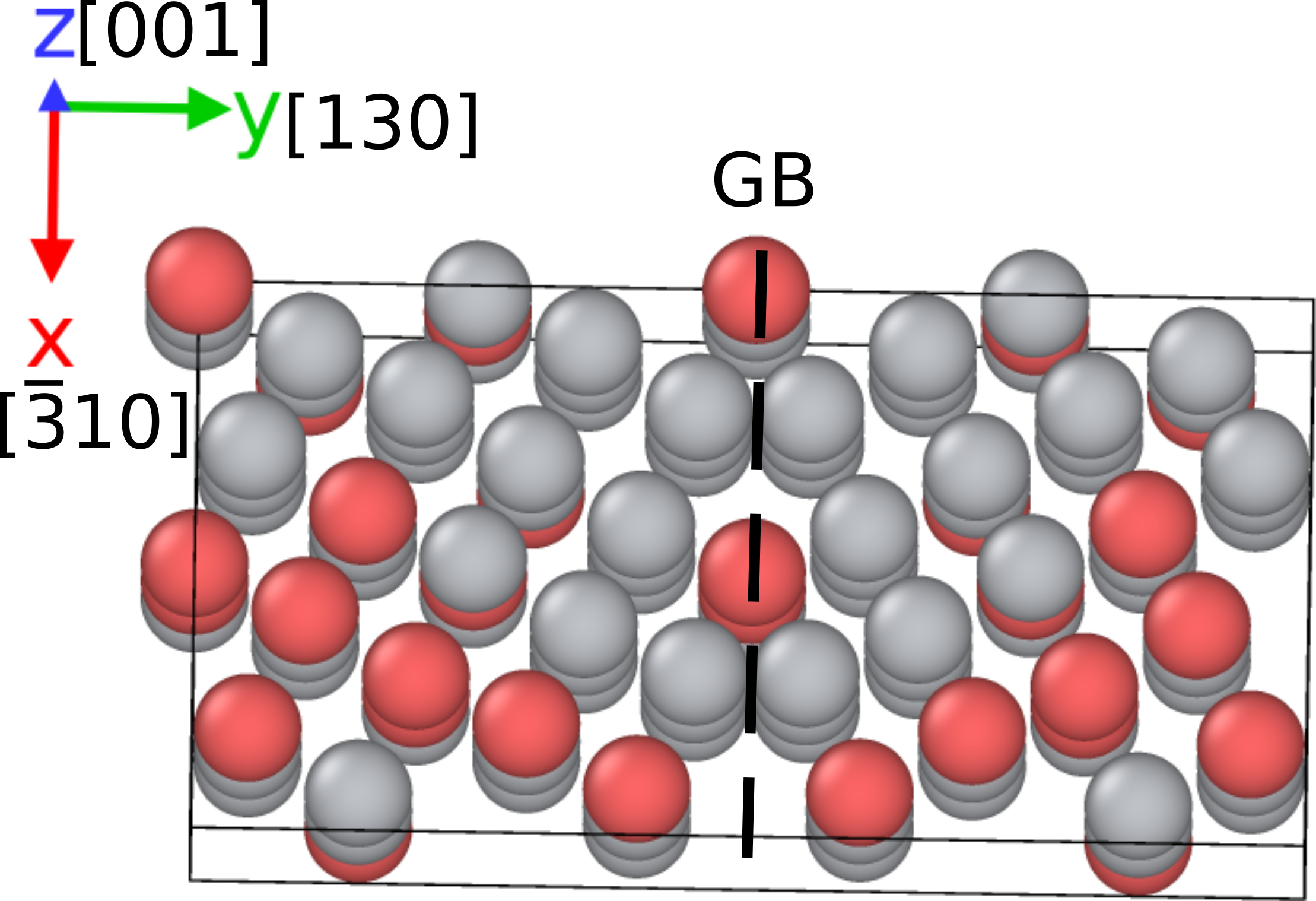}}~~~~~~~\subfloat[\label{fig:gb-2}]{\centering{}\includegraphics[scale=0.3]{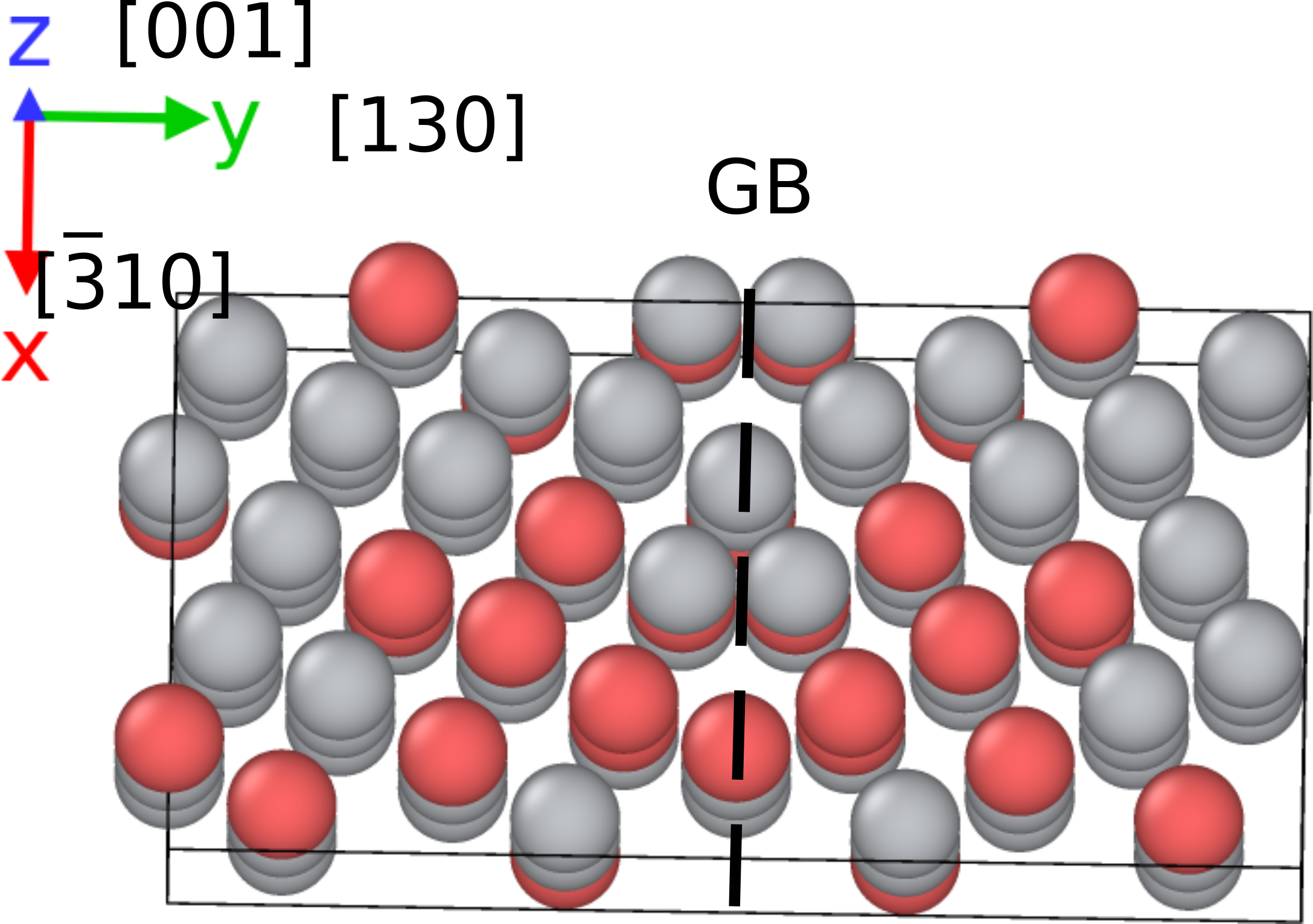}}
\par\end{centering}
\caption{Schematic of all DFT supercells used in the study, grey and red atoms
refer to Ti and Mo sites, respectively. (a) 4x4x4 bulk supercell of
128 atom alloy structure (b) GB-I and (c) GB-II alloy structure}
\end{figure}
\par\end{center}

\subsection{First-principles calculations}

%\begin{doublespace}
\noindent DFT calculations are conducted using the Vienna \emph{ab
initio} simulation package (VASP) with the projector augmented wave
(PAW) pseudopotentials \citep{Blochl1994Dec,Kresse1999Jan}, and the
Perdew-Burke-Ernzerhof (PBE) generalized gradient approximation (GGA)
\citep{Perdew1996Oct} for the exchange-correlation functional to
arrive at the ground-state structures and energies. To truncate the
plane-wave expansion, 360 eV energy cutoff is adopted. A k-mesh density
of 0.02 \AA$^{-1}$ is employed to sample the Brillouin zone.
The outermost electron shell (and number of electrons) in the pseudo
potentials used for Ti, Mo, Y, Zr and Nb are 4s (12), 5s (14), 5s
(11), 5s (12), 5s (13), respectively. A force cutoff of 0.01 eV \AA$^{-1}$ is
adopted for the positions only relaxation step and an energy cutoff
of $10^{-8}$ eV is used for the volume only relaxation step. Periodic
boundary conditions without vacuum are adopted since otherwise the
free surfaces cause instability in the simulations. The Birch-Murnaghan
equation of state (EoS) method \citep{Birch1947Jun} is employed to
relax the GB and GB-solute structures. It is often found that varying
the supercell length perpendicular (i.e., y-direction as shown in
\Figref{gb-1}) to the GB plane changes the excess volume of the GB,
$\delta V$, and thus the GB energy \citep{Scheiber2016Mar}. In our
study, the length of the supercel normal to the plane is varied in steps
of 0.2 \AA ~to change the volume. Four such calculations are
performed in order to obtain the minimum energy for each GB and this
GB structure is then used for any further calculations. Finally, solute
substitution is only done at the GB habit plane in order to preserve
symmetry. For the post-processing step, modules from pymatgen \citep{Ong2013Feb}
and pyiron \citep{pyiron-paper} codebases are used to perform the
EoS fits and obtain the relevant information for Voronoi volume and
local chemistry at each site.
%\end{doublespace}

\section{Results}

\subsection{Solute substitution in bulk}

%\linespread{2}
%\begin{doublespace}
The lattice constant of the Ti-25 at.\% Mo alloy is found to be 3.213$\lyxmathsym{\AA}$. All 128 sites in the alloy bulk structure are
substituted by a third species, i.e., the solute. There are two host
species (Ti and Mo) substitutions for the solute. Since each of the
sites has a different local environment, there is a distribution in
the solute solution energy. In order to compute bulk solution energies,
we first calculate the energy of one Mo atom ($\mathrm{E}_{\mathrm{Mo,alloy}}$)
and one Ti atom ($\mathrm{E}_{\mathrm{Ti,alloy}}$) in the Ti-25 at.\%
Mo alloy to obtain the chemical potential of the two species using:
%\end{doublespace}
\begin{center}
\begin{equation}
\mathrm{E}_{\mathrm{Mo,alloy}}=[\mathrm{E_{125}}(93,32,0)-\unit{\frac{31}{32}}\cdot\mathrm{E_{128}}(96,32,0)]\label{eq:E_Mo_alloy}
\end{equation}
\par\end{center}

\begin{center}
\begin{equation}
\mathrm{E}_{\mathrm{Ti,alloy}}=\unit{\frac{1}{3}}\cdot[\mathrm{E_{128}}(96,32,0)-\mathrm{E_{125}}(93,32,0)]\label{eq:E_Ti_ally}
\end{equation}
\par\end{center}

%\begin{doublespace}
\noindent In the above notation, the number of atoms of each species
is given inside the parenthesis. The first number refers to the majority
species, Ti, the next two numbers refers to the count of Mo atoms
and the solute species, respectively. Bulk solution energy of solute
X at a Ti site is defined as follows:
%\end{doublespace}

\begin{equation}
\mathrm{E_{\mathrm{X@Ti}}^{sol}=E_{128}(95,32,1)-E_{128}(96,32,0)-E_{X,ref}+E_{Ti,alloy}}\label{eq:solution_e_bulk}
\end{equation}
Similarly, for substitution of solute, X at a Mo site one has:
\begin{equation}
\mathrm{E_{\mathrm{X@Mo}}^{sol}=E_{128}(96,31,1)-E_{128}(96,32,0)-E_{X,ref}+E_{Mo,alloy}}\label{eq:solution_e_bulk-1}
\end{equation}
 $\mathrm{E_{X,ref}}$ refers to the energy of a single solute atom
in their ground state structure. For Y and Zr, the hcp structure and
for Nb, the bcc crystal structure is relaxed to obtain the respective
per-atom solute reference energies. 
%\begin{doublespace}
\begin{center}
\begin{figure}[H]
\begin{centering}
\includegraphics[scale=0.9]{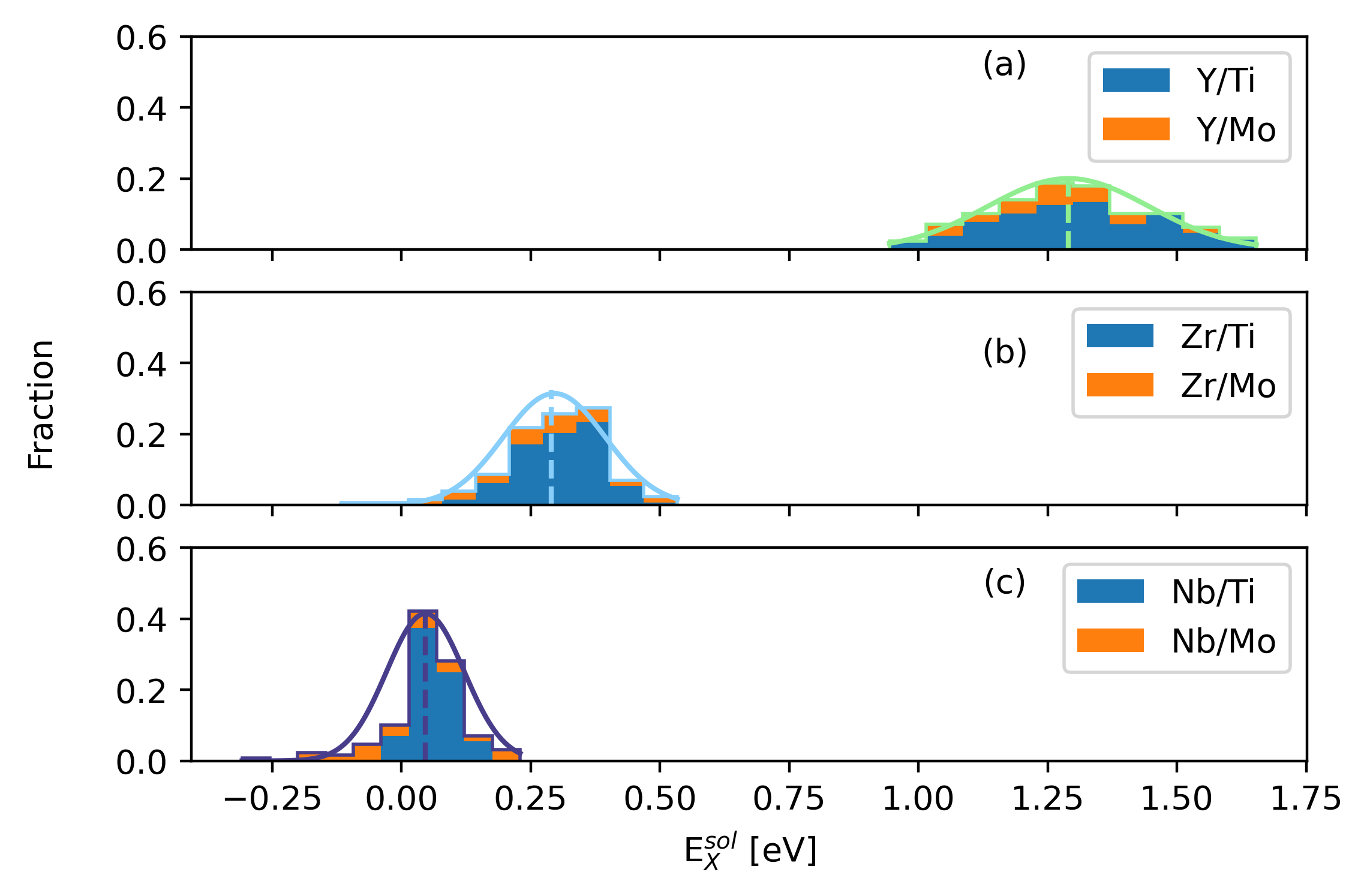}
\par\end{centering}
\caption{DFT calculated bulk solution energy distributions (out of 128 data
points) of (a) Y, (b) Zr and (c) Nb in a Ti- 25at.\% Mo alloy. Solution
energy distributions of each solute is split into Ti and Mo site distributions
indicated by blue and orange bars, respectively. The vertical dashed
lines represent the mean value of the individual distributions. \label{fig:DFT-calculated-bulk}}
\end{figure}
\par\end{center}
%\end{doublespace}

%\begin{doublespace}
The two distributions ($\mathrm{E_{\mathrm{X@Ti}}^{sol}}$ and $\mathrm{E_{\mathrm{X@Mo}}^{sol}}$)
can be reconciled into a single distribution $\mathrm{E_{\mathrm{X}}^{sol}}$
for each solute as shown in \Figref{DFT-calculated-bulk} using the
approach given by \eqref{solution_e_bulk} and \eqref{solution_e_bulk-1}).
These distributions can be described with normal distributions characterized
by a mean value and standard deviation. Please note, in the case of
pure systems, these distributions collapse into a single value or
a Dirac delta distribution. The atomic sites which are at the mean
value are considered as ``representative sites'' for the solutes
in the bulk. From \Figref{DFT-calculated-bulk}, we find that Y (mean:
1.29 eV, std: 0.16 eV) has a strong tendency to not dissolve in the
bulk followed by Zr (mean: 0.3 eV, std: 0.1 eV). This can be rationalized
based on the size of solute atoms, Y being a larger solute atom has
the least tendency to dissolve in the bulk. The range for Y solution
energies is 1.0 - 1.7 eV and for Zr, the range of solution energies
is 0.1 - 0.5 eV. On the other hand, Nb, has comparatively minor solution
energies with a mean of 0.03 eV and a standard deviation of 0.08 eV.
Favourable dissolution, i.e. $\mathrm{E_{sol}^{X}}$ < 0 occurs for
Nb when replacing Mo atoms in the original bulk alloy as illustrated
in \Figref{DFT-calculated-bulk} (c). Although there exists a single
distribution, minor differences between the individual distributions
for Ti and Mo host substitutions are found. For Y, Zr and Nb, the
difference between $<\mathrm{E_{\mathrm{X@Ti}}^{sol}}$> - $<\mathrm{E_{\mathrm{X@Mo}}^{sol}}$>
are 45, 48 and 56 meV, respectively using the mean values of the distributions
shown above. Representative bulk sites are identified when the solute
substitutes at a Ti site and Mo site, respectively. The solution energies
for these representative Ti and Mo sites ($\mathrm{E_{rep}(95,32,1)}$
and $\mathrm{E_{rep}(96,31,1)}$) are 1.33, 0.29, 0.04 eV and 1.28,
0.31, 0.03 eV for Y, Zr and Nb, respectively.
%\end{doublespace}

\subsection{Grain boundary energy of different GB configurations}

%\begin{doublespace}
When creating the Ti-25 at.\% Mo alloy GB structure, there are many
possibilities on mirroring across the y-plane in the bulk supercell.
This leads to different arrangements of Ti and Mo sites in the GB
structure and as a result there are varying concentration of Mo at
the GB. There are ten possible GBs in the present case but only five
of them have stable configurations. The other five configurations
contain less than 20 at.\% Mo in either the bulk or the GB region
of the supercell and lead to significant reconstruction. 

In order to characterize the concentration of Mo at the GB, the structural
unit model of a $\Sigma5$ alloy GB is used, as highlighted in \Figref{bcc-gb-atoms}.
From the schematic, it is clear that the structural repeat units (5
atoms for each unit in $\Sigma5$) can have different arrangement
of Mo atoms in them. To quantify the concentration of Mo in the GB
region, the ratio of number of Mo (small-sized) atoms to the number
of GB (shown in grey) atoms is used.
%\end{doublespace}
\begin{center}
\begin{figure}[H]
\centering{}\subfloat[]{\includegraphics[scale=0.2]{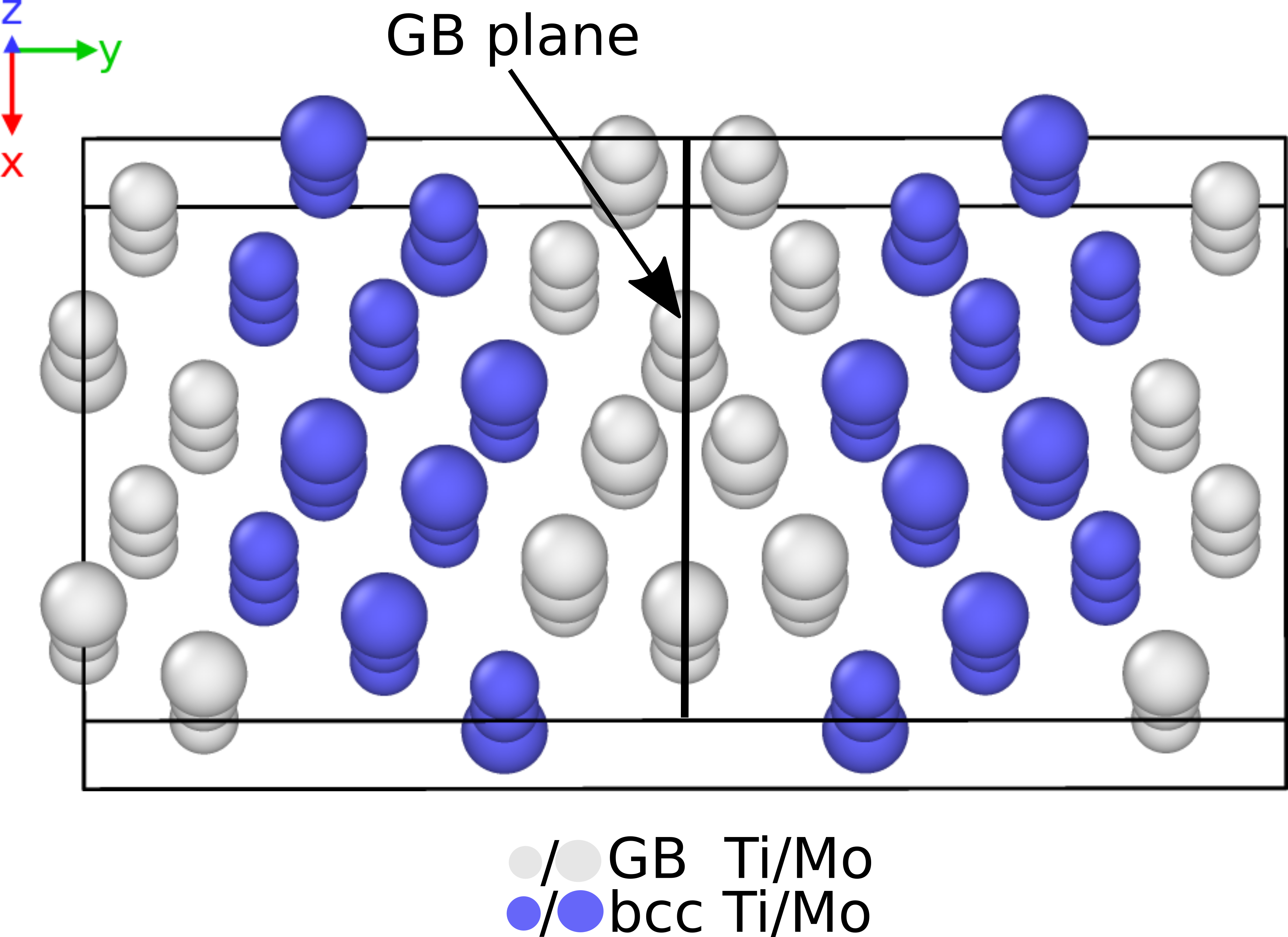}}~~~~~~\subfloat[]{\includegraphics[scale=0.23]{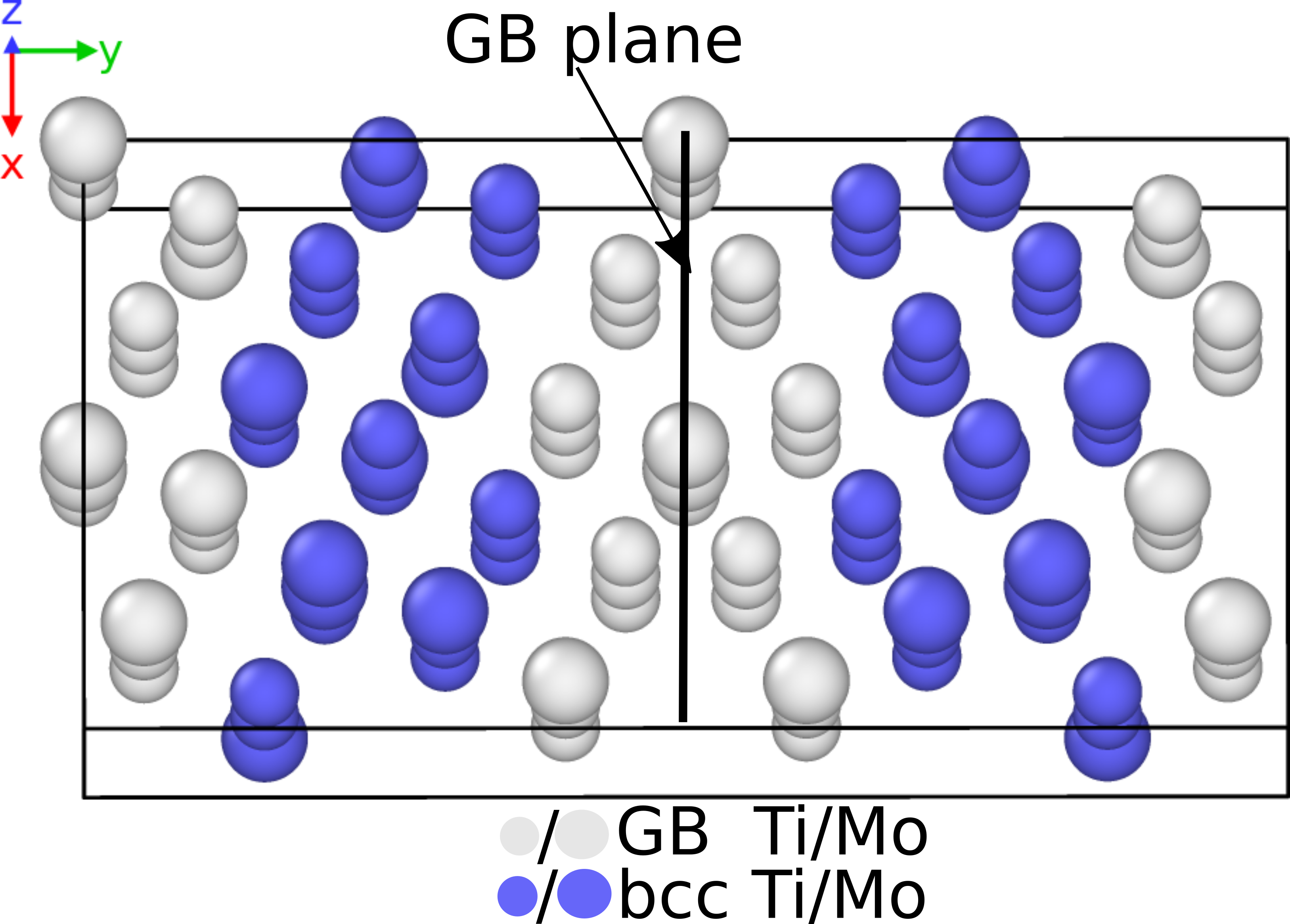}}\caption{Schematic showing the structural units of $\Sigma5$(130){[}001{]}
GB-1 and GB-2 structure in (a) and (b), respectively. GB atoms are
shown in grey and bulk bcc atoms in blue. Ti and Mo species are shown
by small and large atoms, respectively.\label{fig:bcc-gb-atoms}}
\end{figure}
\par\end{center}

%\begin{doublespace}
The GB energy $(\gamma_{\mathrm{GB}})$ is computed as follows:
\begin{equation}
\gamma_{\mathrm{GB}}=[\mathrm{E_{GB}(90,30,0)-90\cdot E_{Ti,alloy}-30\cdot E_{Mo,alloy}]/(2\cdot A_{GB}})\label{eq:gb_e}
\end{equation}
Here, $\mathrm{A_{GB}}$ refers to the area of the GB, and a factor
of two is introduced to account for two GBs since periodic boundary
conditions are used. $\mathrm{E_{GB}(90,30,0)}$ refers to the energy
of the Ti- 25at.\% Mo GB supercell, which varies according to individual
arrangements of Ti and Mo atoms in the GB structure. \Tabref{GB-energy-data}
summarizes the GB energy together with the chemical compositions of
the GB unit cells as well as the GB planes.

\begin{table}[H]
\caption{GB energy-concentration data for five stable configurations \label{tab:GB-energy-data}}

\begin{centering}
\begin{tabular}{c>{\centering}m{4cm}>{\centering}m{4cm}>{\centering}m{4cm}}
\toprule 
S.No & at.\% Mo in GB unit cell (out of 60 atoms) & at.\% Mo at GB plane (out of 12 atoms) & GB energy, $\gamma_{GB}$ \\ (mJ m$^{-2}$)\tabularnewline
\midrule
\midrule 
1 & 20 & 16.6 & 895\tabularnewline
\midrule 
2 (GB-I) & 23.3 & 33.3 & 1100\tabularnewline
\midrule 
3 (GB-II) & 23.3 & 50 & 994\tabularnewline
\midrule 
4 & 26.7 & 33.3 & 1083\tabularnewline
\midrule 
5 & 30 & 33.3 & 1011\tabularnewline
\bottomrule
\end{tabular}
\par\end{centering}
\end{table}

As shown in \Tabref{GB-energy-data}, the GB energies are found to
be approximately 1 J $\mathrm{m}^{-2}$, with a 10\% variation from
0.9 J $\mathrm{m}^{-2}$ to 1.1 J $\mathrm{m}^{-2}$. Within this
range of variation, there is no clear trend with the Mo content in
the GB plane or GB unit cell. Nevertheless, the lowest GB energy is
for the structure with the lowest Mo concentration at the GB unit
cell and GB plane. Similarly, Shen \emph{et al. }\citep{Shen2021Mar}
predicted a non-linear variation of GB energy in bcc U as function
of Mo or Nb content. Further, the role of domain size has been evaluated
by conducting a simulation with a 240-atom cell for the GB-II configuration.
In this case, the calculated GB energy is 966 mJ m$^{-2}$, i.e. within
30 mJ m$^{-2}$ of the GB energy of the 120 atom cell such that it
is concluded that this smaller domain size is sufficient for convergence
of the GB simulations.
%\end{doublespace}

\subsection{GB Segregation in Ti-Mo alloy \label{subsec:Solute-GB-binding-energy}}

\subsubsection{Segregation energies of Ti/Mo to the GB plane}

%\begin{doublespace}
\noindent The two GBs with 23.3 at.\% Mo in the GB unit cell which
is close to the alloy composition are selected for solute substitution
in the GB habit plane. Based on the local GB site chemistries, there
are six different segregation sites for each species, i.e. 3 sites
for Ti and Mo each at GB-I and 4 and 2 sites for Ti and Mo sites at
GB-II, respectively. First, the segregation energies of Mo and Ti
atoms in the binary Ti-Mo system to the GB plane are calculated as
follows: 

\noindent 
\begin{equation}
\mathrm{E_{Mo@Ti}^{seg}=[E_{GB}(89,31,0)-E_{GB}(90,30,0)]-[E_{rep}(95,33,0)-E_{128}(96,32,0)]}\label{eq:host_Mo_bind_e}
\end{equation}

\noindent 
\begin{equation}
\mathrm{E_{Ti@Mo}^{seg}=[E_{GB}(91,29,0)-E_{GB}(90,30,0)]-[E_{rep}(97,31,0)-E_{128}(96,32,0)]}\label{eq:host_Ti_bind_e}
\end{equation}

\noindent In order to calculate the two quantities above, we need
a representative site for a Mo atom at a Ti site in the bulk alloy
and similarly for a Ti atom at a Mo site. We take the average of 96
data points for Mo substitution to be $\mathrm{E_{rep}(95,33,0;bulk)}$
and the average of 32 data points for Mo substitution to be $\mathrm{E_{rep}(97,31,0;bulk).}$
For the GBs, all sites at the GB plane are substituted for Ti and
Mo at Mo and Ti sites, respectively and the results are included in \Figref{DFT-computed-binding-2}. In terms of
statistics, we have 7 sites from GB-I (3 sites) and GB-II (4 sites)
available for Mo substitution. Similarly, we have 5 sites from GB-I
(3 sites) and GB-II (2 sites) available for Ti substitution. Segregation
energies of Ti have a mean of -0.33 eV and range from -0.44 to -0.23
eV. On the other hand, Mo segregation has a mean of 0.44 eV and a
rather large range from 0.1 to 0.69 eV. These results indicate that
Ti has a tendency to segregate to the GBs whereas Mo tends to de-segregate.
%\end{doublespace}

\subsubsection{Solute segregation energies to the GB plane for GB-I and GB-II }

%\begin{doublespace}
\noindent The segregation energy of a solute, X to a Ti site is calculated
as follows:
\begin{equation}
\mathrm{E_{\mathrm{X@Ti}}^{seg}=[E_{GB}(89,30,1)-E_{GB}(90,30,0)]-[E_{rep}(95,32,1)-E_{128}(96,32,0)]}\label{eq:Ti_bindGB}
\end{equation}

%\end{doublespace}

%\begin{doublespace}
Similarly, for a Mo site, the segregation energy is computed as: 
%\end{doublespace}

\begin{equation}
\mathrm{E_{\mathrm{X@Mo}}^{seg}=[E_{GB}(90,29,1)-E_{GB}(90,30,0)]-[E_{rep}(96,31,1)-E_{128}(96,32,0)]}\label{Mo_bind_GB}
\end{equation}

%\begin{doublespace}
\Figref{DFT-computed-binding-2} shows the segregation energies of
the three solutes in GB-I and GB-II. For Y, there is a case when substituting
Ti in GB-II leads to a structural change of the GB resulting in a
large (-2.02 eV) segregation energy which is not shown in \Figref{DFT-computed-binding-2}.
Overall, there is a clear trend for the segregation energies of the
three solutes. Y has the strongest attraction to the GBs followed
by Zr whereas Nb shows only comparatively weak attraction when replacing
Mo but not when replacing Ti. This is consistent with their bulk misfit
as shown above based on solution energies. Further, for all solutes
replacing Mo in the GB is preferred as compared to Ti substitution
which is consistent with the segregation energies of Mo and Ti at
the GB sites in the binary Ti-Mo system, i.e. there is a preference
of Ti segregation and de-segregation of Mo.
%\end{doublespace}
\begin{center}
\begin{figure}[H]
\includegraphics[scale=0.7]{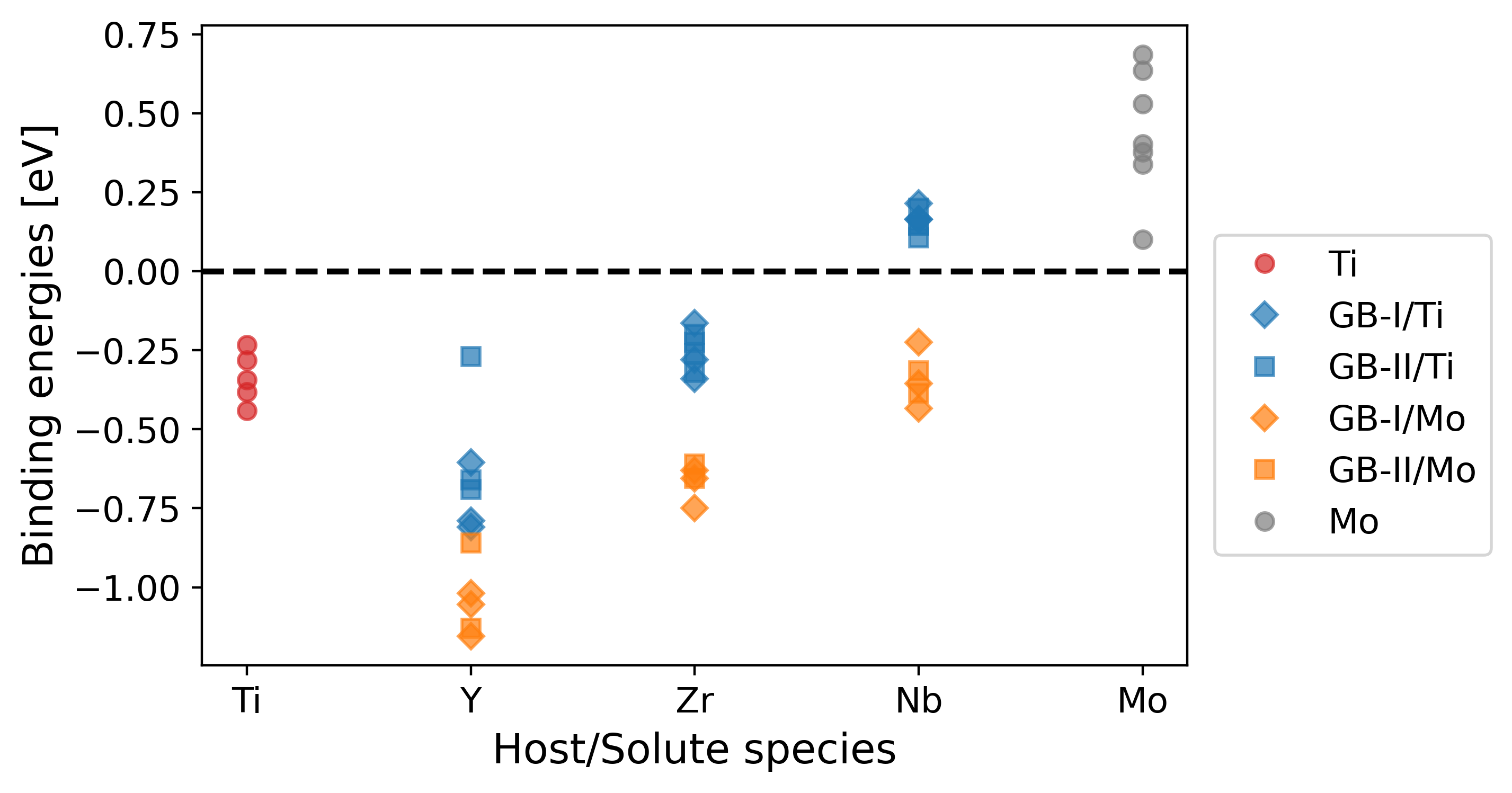}
\centering{}\caption{\label{fig:DFT-computed-binding-2}DFT computed segregation energies
of Y, Zr and Nb to Ti and Mo host sites in GB-I (diamond markers)
and GB-II (square markers). Individual points are colored based on
whether Ti (blue) or Mo (orange) is substituted at the GB plane. Segregation
energies below the dashed line suggest favorable segregation, while
above the dashed line suggest unfavorable segregation.}
\end{figure}
\par\end{center}

\section{Discussion}

\subsection{Bulk properties}

\begin{center}
\begin{figure}[H]
\noindent \centering{}\includegraphics[scale=0.6]{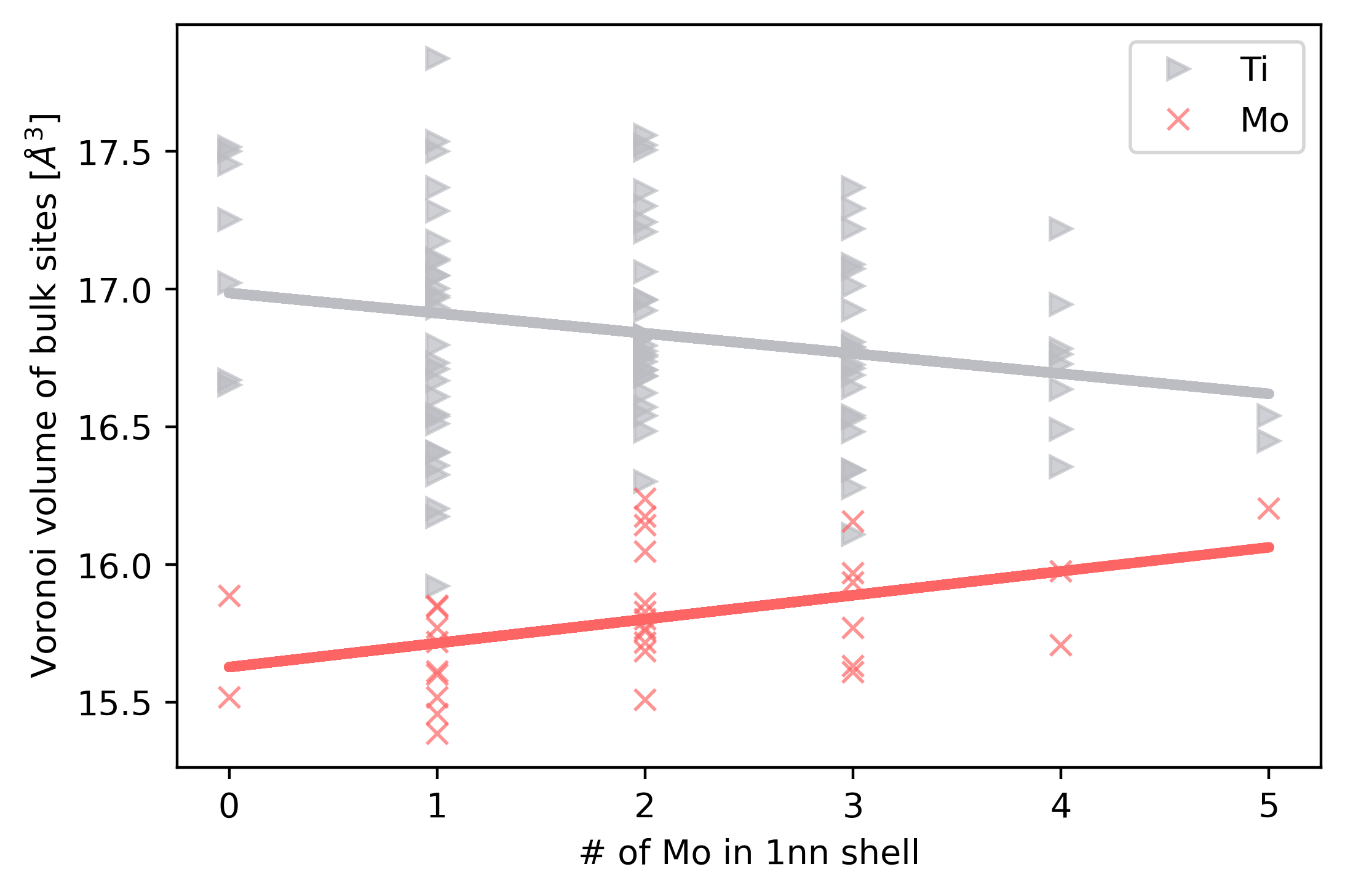}\caption{\label{fig:voronoi_at_Mo_1nn_Mo_bulk}Voronoi volume of Ti and Mo
sites in the bulk alloy as a function of number of Mo atoms in the
1nn shell (out of 8 atoms).}
\end{figure}
\end{center}

\begin{figure}[H]
\begin{centering}
\subfloat[]{\includegraphics[scale=0.6]{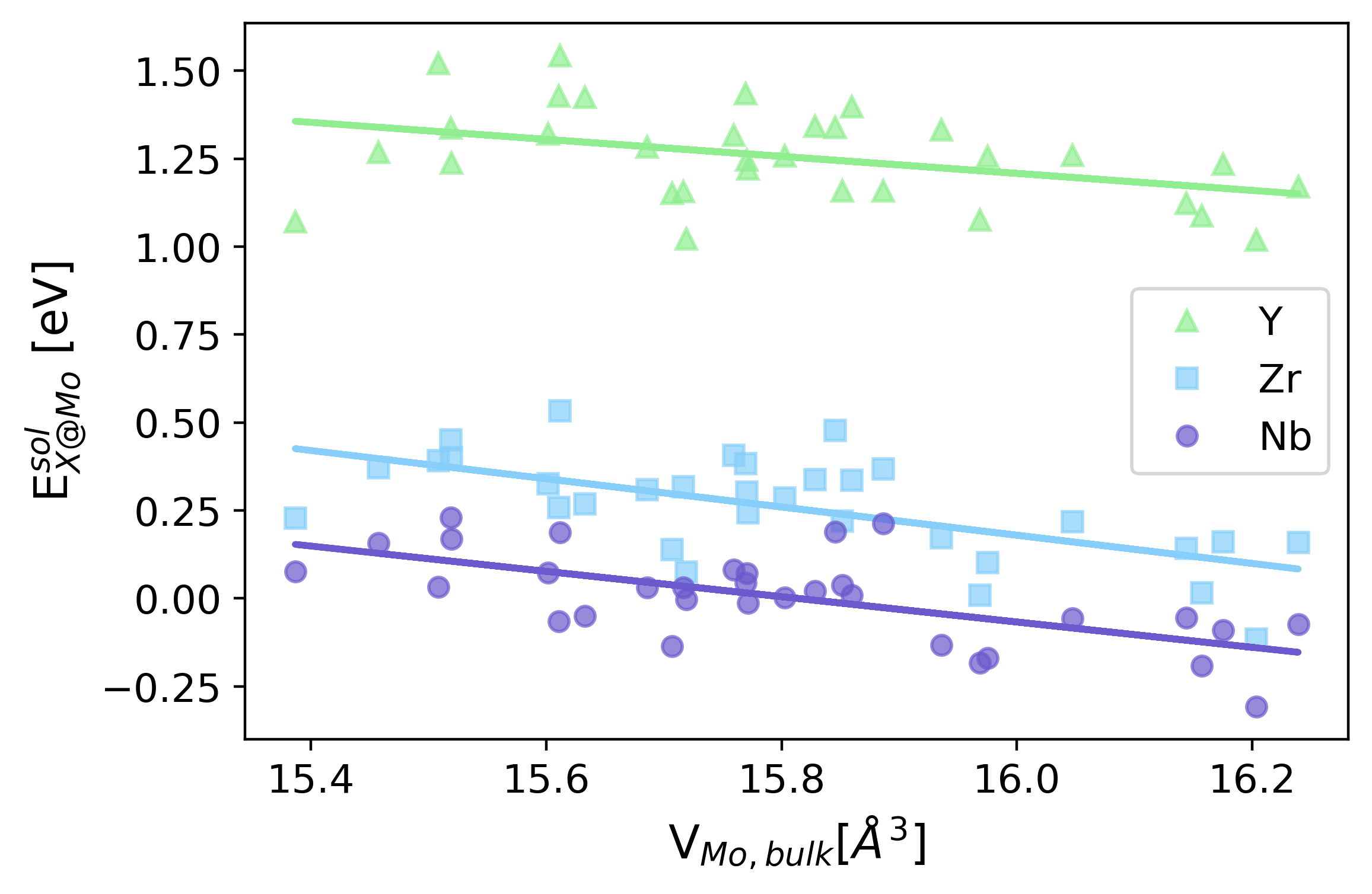}}\subfloat[]{\includegraphics[scale=0.6]{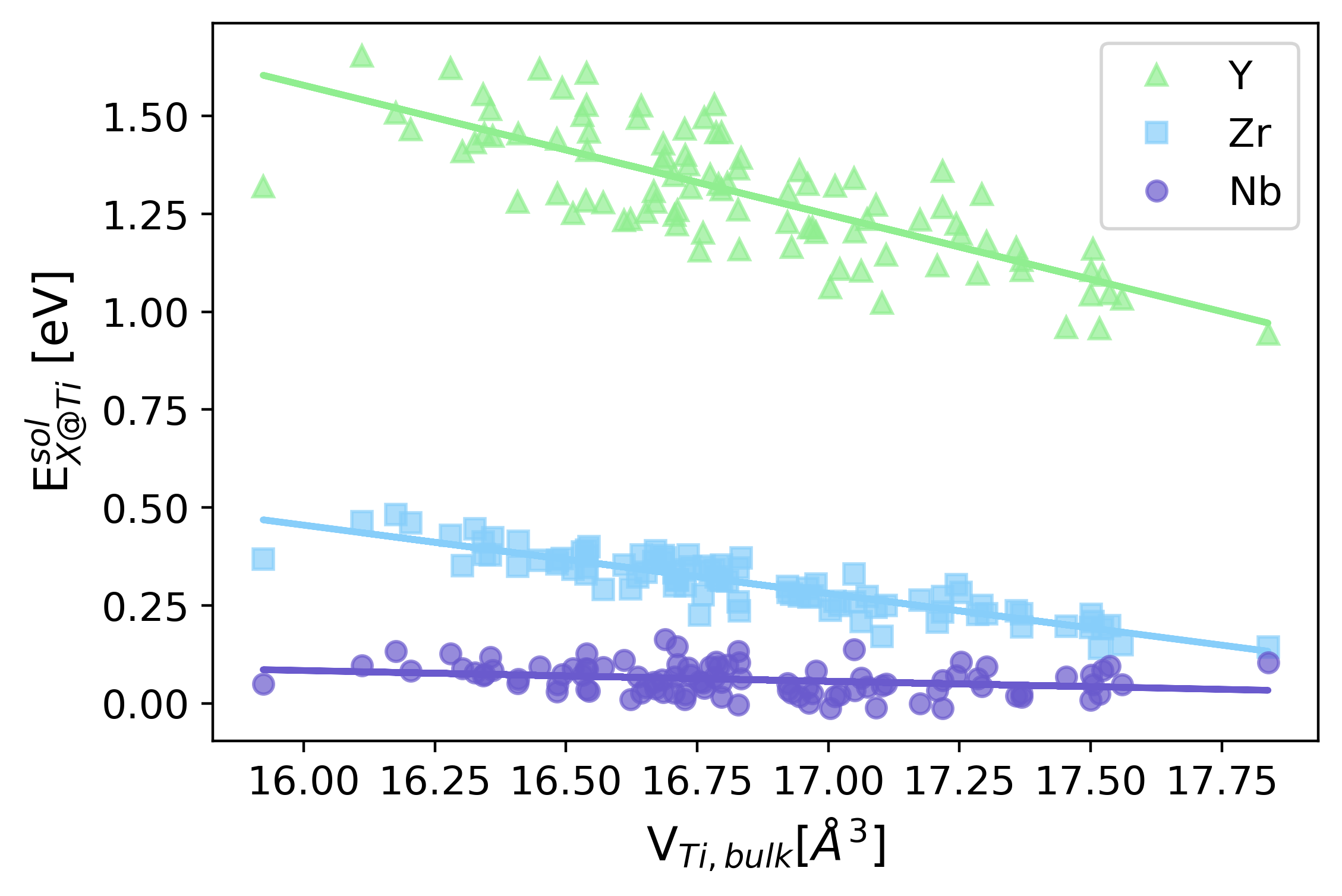}

}
\end{centering}
\caption{(a) Computed bulk solution energy at a Mo site plotted as a function
of Voronoi volume of the same Mo site before solute substitution.
(b) Computed bulk solution energy at a Ti site plotted as a function
of Voronoi volume of the same Ti site before solute substitution
\label{fig:bulk-voronoi-e-sol}}
\end{figure}

%\begin{doublespace}
\noindent In the bulk alloy, there 128 sites of Ti and Mo in total.
The Voronoi volume of these sites is shown in \Figref{voronoi_at_Mo_1nn_Mo_bulk}
as a function of the number of Mo atoms in the first next neighbor
(1nn) shell of each atomic site. With an increasing number of Mo atoms
in the 1nn shell, the local Voronoi volume decreases for Ti and increases
for Mo sites, respectively. On average, there is a linear trend in
both cases. Further, the local volume of Ti sites is in general larger
than that of Mo sites.
%\end{doublespace}
\noindent 
%\begin{doublespace}
\noindent In order to decouple the local chemistry and local volume
effects on the bulk solution energies, the trends between solution
energies and Voronoi volume before solute substitution is investigated.
As shown in \Figref{bulk-voronoi-e-sol}, the trends for Y, Zr and
Nb's bulk solution energy are the same except for Nb on Ti sites where
the solution energies are negligible. In general there is a linear
decrease of solution energies with increasing site volume. This trend
is particularly pronounced for Y when replacing Ti.
%\end{doublespace}
\\
%\begin{doublespace}
\noindent The negative slope can be rationalized by size effects,
Y being a larger solute atom prefers to dissolve in sites which are
larger in size in the distribution of available sites. Similarly,
Zr has a moderate linear dependence with the Voronoi volume indicating
the size effects persists although to a smaller extent than for Y.
In the case of Nb, the dependence with the volume is negligible and
this can be rationalised by the fact that Nb is a smaller solute compared
to the other two solutes.
%\end{doublespace}
%\begin{doublespace}
\noindent In addition to site volume, the local neighborhood can affect
the bulk solution energy distribution. To analyze the potential chemistry
effect the number of Mo/Ti atoms in the 1nn shell may be considered
as a measure. The results of this analysis are shown in \figref{1nn-analysis}.
As noted in ref. \citep{Ruban2016Apr}, the slopes of the trend lines
are essentially the interaction parameters between the solute and
the host atom that the solute is being substituted for. Slopes of
the lines of Nb (-0.1 eV/at) and Zr (-0.09 eV/at) substituting a Mo
site indicate towards to favourable dissolution when increasing the
number of Mo atoms in the local chemical neighborhood. This could
be because of chemical interaction (i.e., electronegativity) between
Nb and Mo as found in \citep{Rojas2022Feb} when studying high-entropy
alloys containing Mo and Nb. Another key observation is the dissolution
of Y in a Ti site (slope of -0.08 eV/at); it becomes favorable to
dissolve a Y atom in a Ti site with local neighborhoods having a higher
number of Ti atoms. This can be rationalized by the strain-relief
argument where Y atom likes to dissolve at the larger volume (Ti)
sites than smaller volume (Mo) sites as seen in \figref{bulk-voronoi-e-sol}(a)
and \figref{bulk-voronoi-e-sol}(b), respectively. This variation
indicates that there might be synergistic effects of local volume
and chemistry for the case of Y at Ti substitution. Finally, there
is almost no correlation between Nb and Zr substitution energies at
Ti sites and the number of Ti in the first neighbor shell. This is
true for Y substitution at Mo sites as well. Thus, the dominant
effect on the bulk solution energies is the local Voronoi volume for these cases. 
%\end{doublespace}

\begin{center}
\begin{figure}[H]
\begin{centering}
\subfloat[]{\includegraphics[scale=0.6]{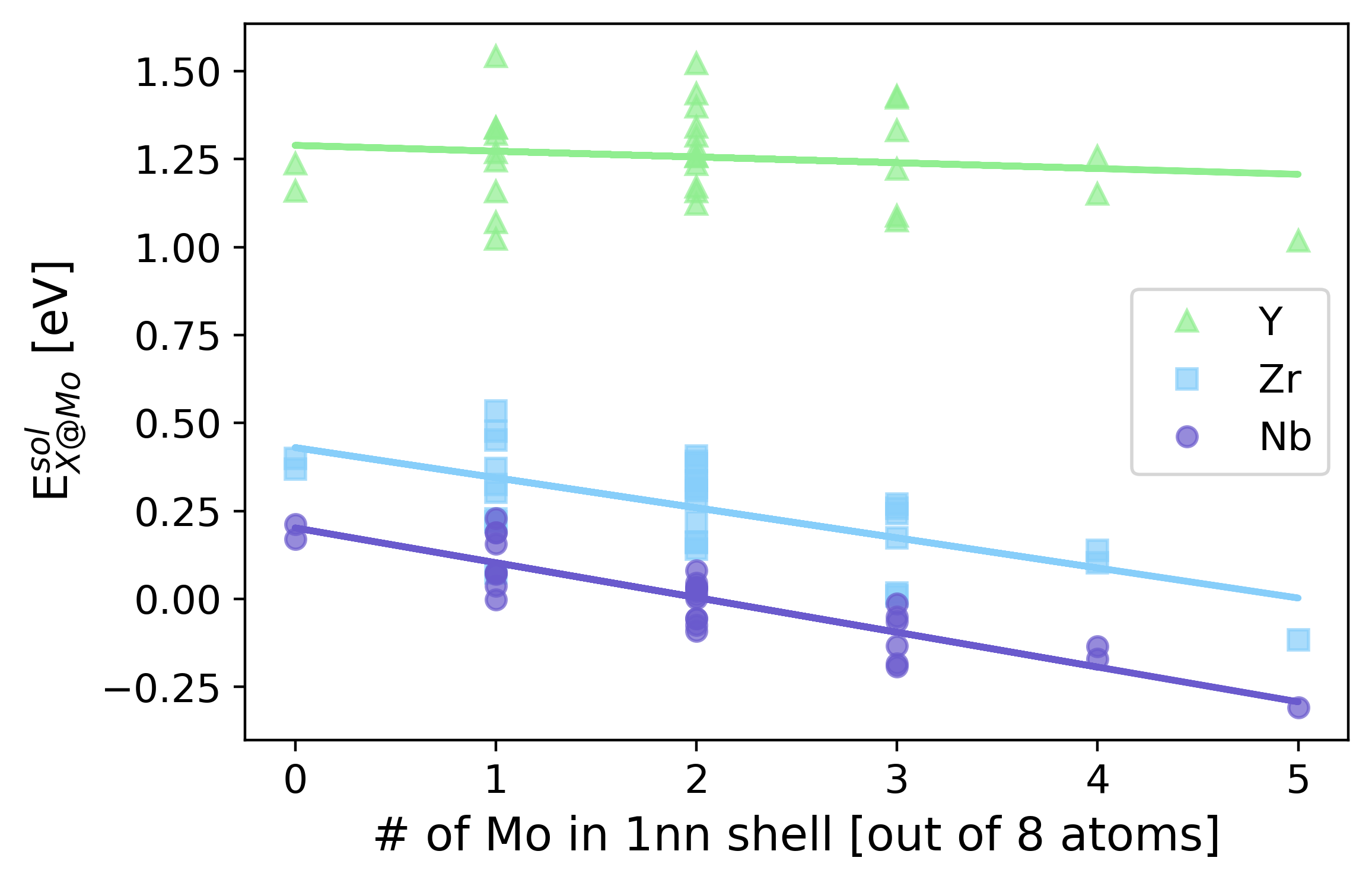}

}\subfloat[]{\includegraphics[scale=0.6]{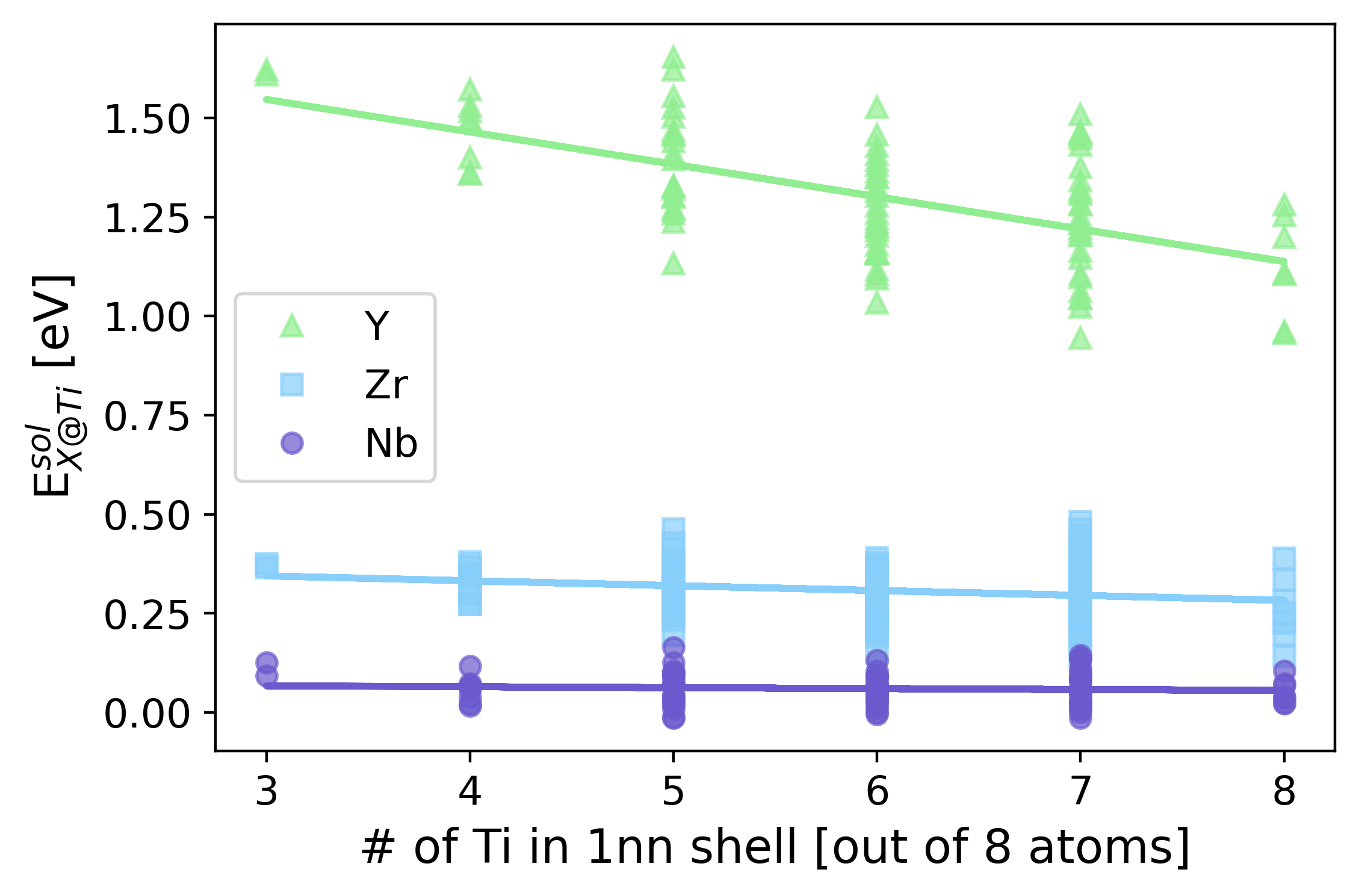}

}
\par\end{centering}
\caption{(a) Computed bulk solution energy at a Mo site plotted as a function
of number of molybdenum atoms in first nearest neighbor (1nn) shell
around each solute-substituted site. (b) Computed bulk solution energy
at a Ti site plotted as a function of number of titanium atoms in
first nearest neighbor (1nn) shell around each solute-substituted
site. The trend lines are obtained by averaging the values at number
of sites having the same \# of Ti/Mo atoms in 1nn shell.\label{fig:1nn-analysis}}
\end{figure}
\par\end{center}

\subsection{\label{subsec:Analysis-of-GB-solute}GB properties}

\begin{figure}
\subfloat[\label{fig:Gb-analysis-voronoi-volume-solute}]{\begin{centering}
\includegraphics[scale=0.6]{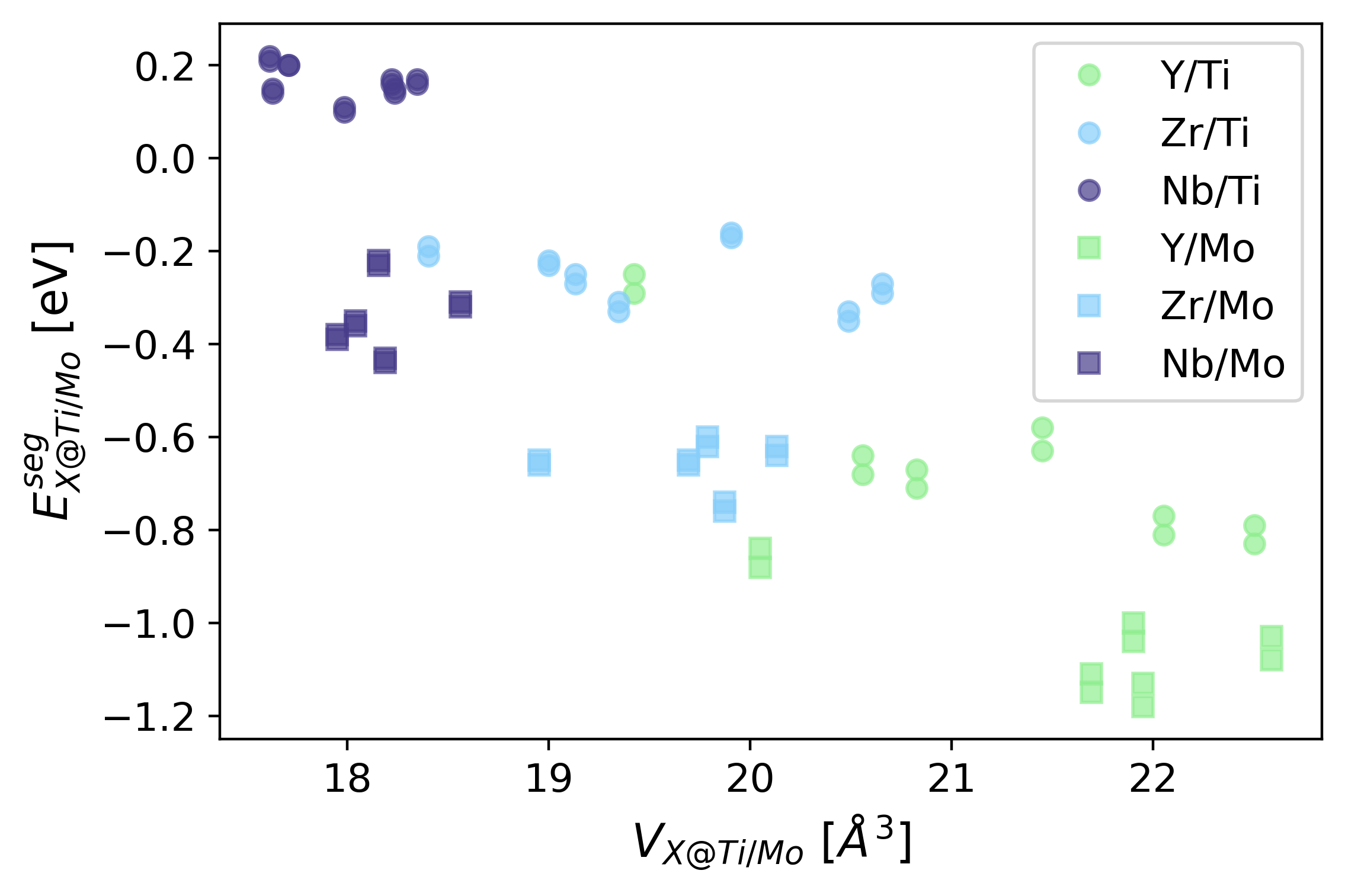}
\par\end{centering}
\centering{}}\subfloat[\label{fig:Gb-analysis-voronoi-volume}]{\begin{centering}
\includegraphics[scale=0.6]{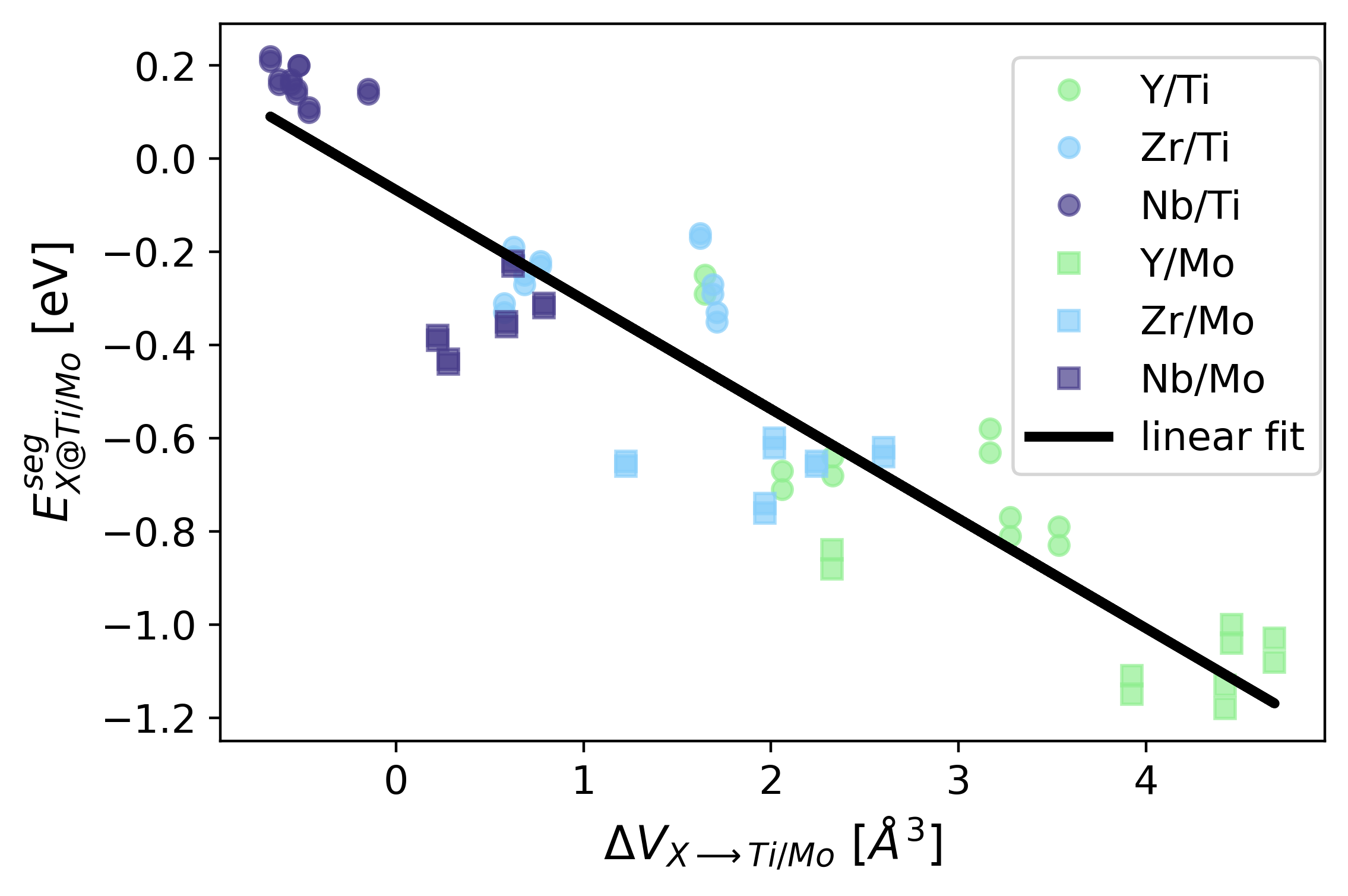}
\par\end{centering}
\centering{}}\caption{(a) DFT computed segregation energies of Y, Zr and Nb to Ti (circles)
and Mo (squares) host sites in GB-I and GB-II as a function of Voronoi
volume of solute substituted sites ($\mathrm{V_{X@Ti/Mo}}$ )(b)DFT
computed segregation energies of Y, Zr and Nb to Ti (circles) and
Mo (squares) host sites in GB-I and GB-II as a function of volume
difference between Voronoi volume after and before solute substitutions
at GB plane sites ($\mathrm{\Delta V_{X->Ti/Mo}}$)}

\end{figure}

%\begin{doublespace}
\noindent The differences in solute segregation to Ti and Mo sites
are analyzed in further detail. First, a discussion on Voronoi volume
of the undecorated and decorated GB site is presented. A trend analysis
of segregation energies with Voronoi volume of clean GB sites shows
that the segregation energies of various solute-host pairs are essentially
independent of the Voronoi volume of clean GB sites. Next, the trends
of solute segregation energies as a function of Voronoi volume of
solute substituted sites is presented as shown in \Figref{Gb-analysis-voronoi-volume-solute}
where at least for Y the magnitude of the segregation energy increases
with site volume. The trends for Zr and Nb remain negligible. Although
the individual data for Ti and Mo substitution seem to have similar
volume ranges, the segregation energies vary for both cases.
%\end{doublespace}

%\begin{doublespace}

%\end{doublespace}

%\begin{doublespace}
\noindent However, the volume difference before and after solute substitution
provides a clearer trend for the resulting segregation energies. This
is shown in \figref{Gb-analysis-voronoi-volume}, where the x-axis
represents the change in site volume between solute segregated site
and that of the clean GB site, i.e. $\mathrm{\Delta V_{X->Ti/Mo}}$=$\mathrm{V_{X,i}-V_{Ti/Mo,i}}$.
Here, $\mathrm{V_{X,i}}$is the Voronoi volume of solute at GB site
i and $\mathrm{V_{Ti/Mo,i}}$ is the Voronoi volume of the Ti/Mo atom
at the same GB site i. A general observation is that a volume increase
favours segregation whereas a decrease in volume leads to a positive,
i.e. unfavourable, segregation energy as is the case for Nb on Ti
sites. Nb segregating to a Ti site occupies a smaller volume than
its undecorated counterpart site which is associated with positive
(unfavourable) segregation energies. Further, a general trend is that
the volume changes for Mo sites are larger than for Ti sites and hence
this further indicates why the magnitude of favourable segregation
energies to a Mo site is larger. This volume difference scales linearly
with the solute segregation energy, where a larger volume difference
implies a stronger solute segregation. 

\noindent A consistent shift of segregation energies when replacing
Mo versus Ti is also predicted for Y, Zr and Nb, i.e. the attraction
for these solutes to the GB is larger when replacing Mo. Excluding
the outlier (-0.25 eV) for Y at Ti host substitution, differences
between $<\mathrm{E_{\mathrm{X@Ti}}^{bind}}>$ and $<\mathrm{E_{\mathrm{X@Mo}}^{bind}>}$
for Y, Zr and Nb are 0.41, 0.41 and 0.50 eV, respectively. However,
these differences do not exactly match with the difference of 0.77
eV between Ti and Mo segregation energies in the Ti-Mo system (see
\Figref{DFT-computed-binding-2}). Thus, there is also a solute specific
role in changing the GB site volume.

\noindent The above analysis indicates that elastic interactions are
significant for the magnitude of the segregation energies. In order
to obtain average segregation energies, the evaluation of an Eshelby-type
approach may be considered using information of the undecorated GB
structures as proposed in previous studies \citep{Huber2017Jun,Geng2001Apr}.
From \Figref{Gb-analysis-voronoi-volume}, it is seen that there is
primarily elastic contribution to the segregation energies of solutes.
Scheiber \emph{et al.} \citep{Scheiber2016Nov} have proposed an expression
which captures these interactions based on an Eshelby-type model \citep{Eshelby1956Jan,Eshelby1954Feb}
:
\begin{equation}
\mathrm{E_{X@Ti/Mo}^{seg,el}=\frac{-(V_{A}-V_{B})^{2}}{(\frac{3}{2}\cdot\frac{V_{A}}{G_{A}}+2\cdot\frac{V_{B}}{K_{B}})}+\frac{(V_{A}^{GB}-V_{B})^{2}}{(\frac{3}{2}\cdot\frac{V_{A}^{GB}}{G_{A}}+2\cdot\frac{V_{B}}{K_{B}})}+\frac{-(V_{A}^{GB}-V_{A})^{2}}{(\frac{3}{2}\cdot\frac{V_{A}^{GB}}{G_{A}}+2\cdot\frac{V_{A}}{K_{A}})}}\label{eq:eshelby_binding_energy}
\end{equation}
Here, $\mathrm{V_{A}}$, $\mathrm{V_{B}}$ and $\mathrm{V_{A}^{GB}}$
refers to bulk volume of host and solute as well as volume of the
GB sites, respectively. The elastic properties, $\mathrm{G_{A}}$,
$\mathrm{K_{A}}$ and $\mathrm{K_{B}}$ are shear modulus of host,
bulk modulus of host and solute, respectively. These properties for
the host atom are taken to be that of Ti-Mo alloys: 100 GPa for Young's
modulus from \citep{Yan2017Jul} and a poisson's ratio of 0.33 was
assumed which allowed to calculate $G_{A}$ and $K_{A}$ to be 98
GPa and 37.6 GPa, respectively. $V_{A}$ is determined based on the
calculated lattice constant for the Ti-Mo alloy of 3.213 \AA.
In the case of Y, Zr and Nb as solutes (B), the properties are taken
from Bakker \citep{Bakker1998}. As proposed by Scheiber \emph{et
al.} a scaling constant $\mathrm{c}$ can be introduced for $V_{A}^{GB}$
to improve fits to the calculated segregation energies by changing
the clean GB site volume by multiplying it with $\mathrm{c}$. Since,
there exists a difference in solute segregation energies when substituting
Ti or Mo sites (see \Figref{DFT-computed-binding-2}), two scaling
factors ($c_{Ti}$ and $c_{Mo}$) are quantified, one for each host
species. The scaling factors $c_{Ti}$ and $c_{Mo}$, are found to
be 1.2 and 1.5, respectively reflecting the larger volume change of
Mo sites after solute substitution. Whilst, these models do not capture
the true complexity of each site they still give overall trends of
solute segregation energy predictions for the considered alloy system.
The predictions from the model based on elastic interactions are summarized
in \Figref{seg-e-dft-vs-eshelby} indicating that the Y and Zr average
segregation energies can be explained well within this framework,
while Nb predictions are a bit off in from the DFT calculated values.
Thus, other factors as e.g. chemical interactions may affect the segregation energies of smaller-sized solutes such as Nb. Still, the deviations are within 0.25~eV and general trends among the solutes are predicted correctly, nonetheless. If reasonable predictions could be made in this
way one would not have the need for performing various solute-GB calculations.
For example, it was observed by Huber \emph{et al.} \citep{Huber2017Jun}
that solute segregation energies of Pb and Mg in Al scale linearly
with Voronoi volume of clean sites in symmetric and general tilt boundaries. 
%\end{doublespace}
\begin{center}
\begin{figure}
\begin{centering}
\includegraphics[scale=0.64]{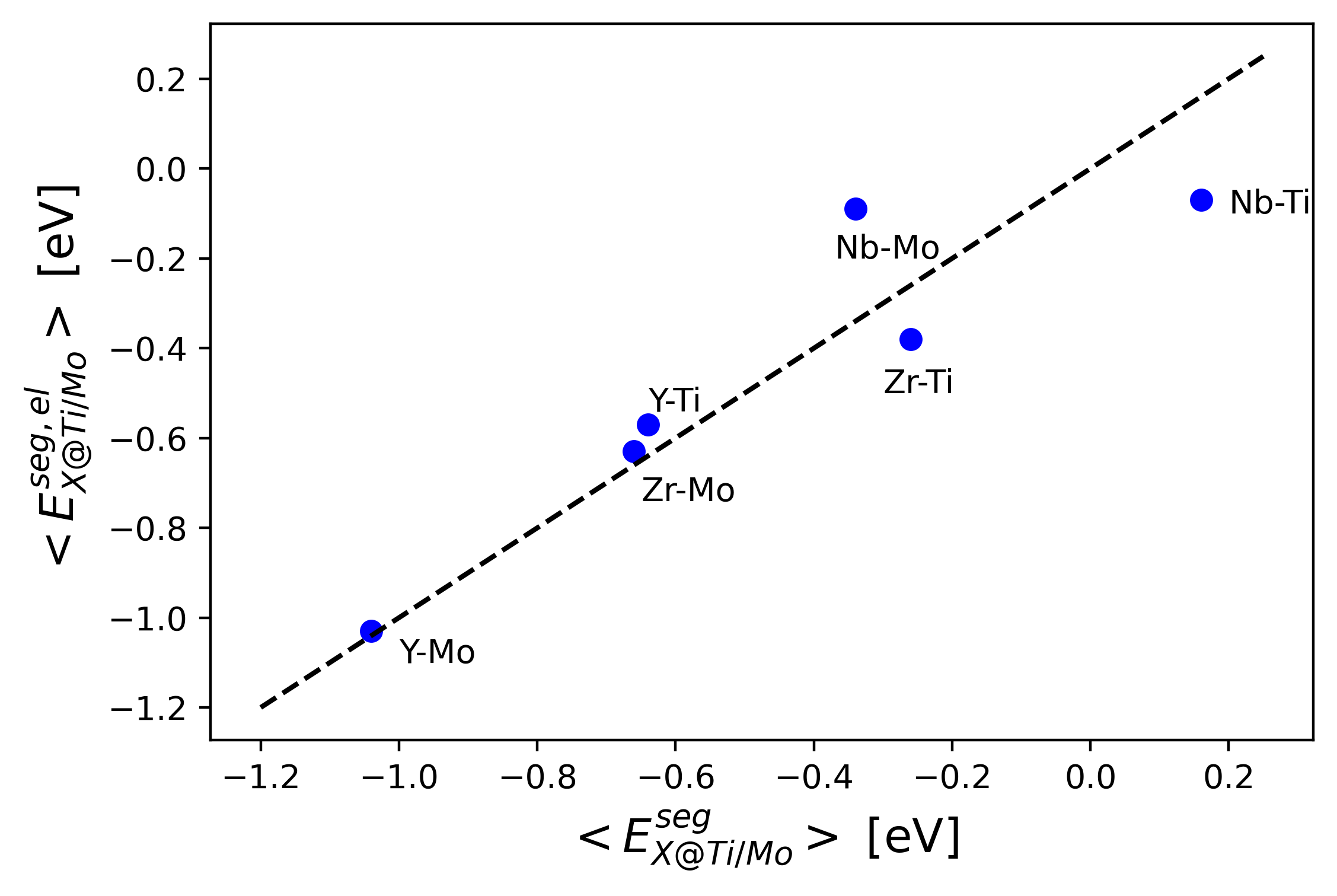}
\par\end{centering}
\caption{\label{fig:seg-e-dft-vs-eshelby}Comparison of average of segregation
energies for Y, Zr and Nb to Ti and Mo sites calculated using DFT
and Eshelby-type model}
\end{figure}
\par\end{center}

%\begin{doublespace}
\noindent Using the above developed Eshelby-type model for two host
sites- Ti and Mo, average GB solute segregation energies for other
transition metal (TM) solutes can be estimated for the Ti- 25 at\%.
Mo alloy. In \Figref{eshelby_TM_solutes} (a), the predicted energies
are presented using the scaling constants obtained above for Ti and Mo sites,
as well as the data for bulk modulus and volume of solutes from Bakker
\citep{Bakker1998}. The observed trends correlate strongly with
the solute volume shown in \Figref{eshelby_TM_solutes}(b). The larger
solutes show similar behaviour as Y and Zr, i.e. they have a tendency
to segregate to the GB with a preference for the Mo sites. On the
other hand, smaller solutes, i.e. those in the middle of the TM series,
have a tendency to de-segregate and the repulsion from the Mo sites
is larger than that from the Ti sites. Overall, the simplified analysis
provides a first guidance regarding the solute-GB interaction for
the given alloy provided elastic contributions are dominant.
%\end{doublespace}
%\begin{doublespace}
\begin{center}
\begin{figure}[H]
\begin{centering}
\includegraphics[scale=0.5]{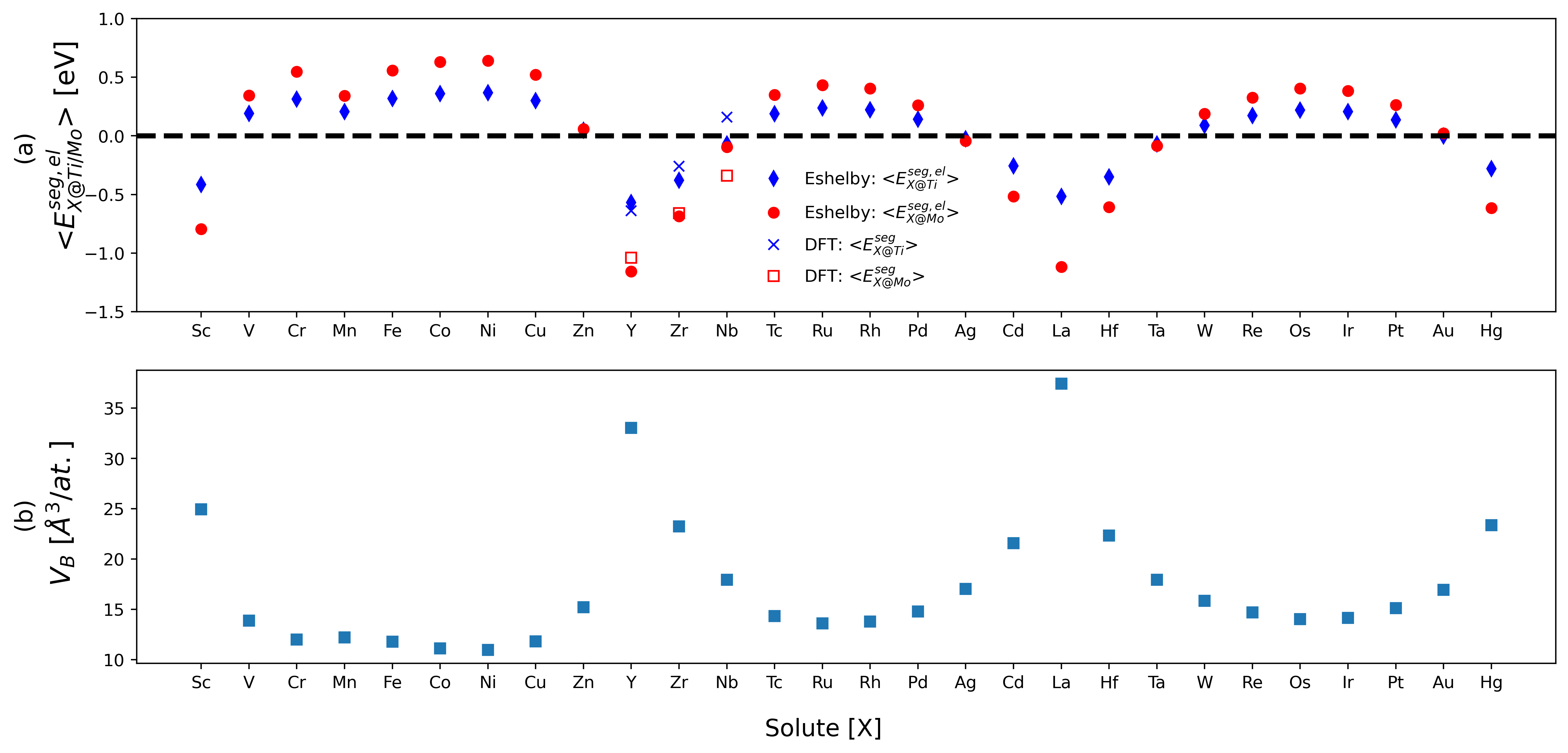}
\par\end{centering}
\caption{(a) Eshelby-type model predictions for segregation energies of (\emph{3d-,
4d-, 5d}-) transition metal solutes to the $\Sigma5$ GB plane in a Ti-Mo alloy system and average segregation energies of Y, Zr and Nb from DFT (b)  Atomic volume of solute elements as taken from Bakker \citep{Bakker1998}\label{fig:eshelby_TM_solutes}}
\end{figure}
\par\end{center}
%\end{doublespace}

\section{Conclusions}

%\begin{doublespace}
This study lays out a first-principles approach for investigating
solute segregation in highly alloyed systems i.e., here Ti-25 at.\%
Mo based on the SQS approach. The main findings of this work are as follows:
%\end{doublespace}
\begin{enumerate}
%\begin{doublespace}
\item A rigorous approach has been applied to determine solution energies
in the bulk by considering all possible permutations in the simulation
domain. The resulting solution energy distributions provide the basis
to identify representative bulk energies for each considered solute.
In the present case, i.e. for Y, Zr and Nb, elastic interactions dominate
such that the volume of the bulk site rather than its local chemistry
determine the mean solution energy which increase with solute size,
i.e. Nb < Zr < Y. In these elastically dominated cases, the identification
of the averaged bulk volume site in the alloy without solutes can
be used to calculate the representative solution energy.
\item The solute segregation energies to GBs is also dominated by elastic
interactions. The situation at the GB is, however, more complicated
as the elastic interaction may be affected by the local chemistry.
In the investigated cases, the change in GB volume sites before and
after solute substitution has been identified as the critical aspect
that determines the solute segregation energy. This volume change
has been found to depend on which of the host species is replaced,
i.e. it is larger for Mo sites than for Ti sites. Thus, the magnitude
of segregation energies are larger when Mo sites are replaced and
overall the segregation energies follow the solute size with Y having
the largest attraction to the GBs followed by Zr and Nb.
\item Using an Eshelby-type approach with a host species dependent scaling
factor for the grain boundary site volume provides a satisfactory
description of the mean segregation energies computed from the DFT
simulations for strongly segregating solutes (here Y, Zr), i.e. solutes
that are significantly larger than the host atoms. The Eshelby-type
analysis has been extended to three different series of transition
metals showing a clear trend of segregation energies with solute size,
i.e. larger solutes are more attracted to smaller host sites (Mo)
and smaller solutes are less repelled from larger host sites (Ti). 

\section*{Acknowledgements}
The authors gratefully acknowledge the financial support under the scope of the COMET program within the K2 Center ''Integrated Computational Material, Process and Product Engineering (IC-MPPE)'' (Project No 886385). This program is supported by the Austrian Federal Ministries for Climate Action, Environment, Energy, Mobility, Innovation and Technology (BMK) and for Labour and Economy (BMAW), represented by the Austrian Research Promotion Agency (FFG), and the federal states of Styria, Upper Austria and Tyrol. This research was funded also in part by the Austrian Science Fund (FWF) (P 34179-N).

%\end{doublespace}
\end{enumerate}
\bibliographystyle{bib/elsarticle-num}
\bibliography{bib/references}

\end{document}